\documentclass[aps,pra,twocolumn,reprint]{revtex4-2}
\usepackage{graphicx}
\usepackage{amsmath, amssymb, amsfonts}
\usepackage{xcolor}
\usepackage[normalem]{ulem}

\begin{document}
\title{Microwave shielding of ultracold polar molecules on the transition $\boldsymbol{n=1 \rightarrow 2}$}

\author{Joy Dutta}
{\email{joy.dutta@durham.ac.uk} \affiliation{Joint Quantum Centre (JQC)
Durham-Newcastle, Department of Chemistry, Durham University, South Road,
Durham, DH1 3LE, United Kingdom.}

\author{Jeremy M. Hutson}
{\email{j.m.hutson@durham.ac.uk} \affiliation{Joint Quantum Centre (JQC)
Durham-Newcastle, Department of Chemistry, Durham University, South Road,
Durham, DH1 3LE, United Kingdom.}

\date{\today}

\begin{abstract}
We show that microwave shielding on the rotational transition $n=1\rightarrow 2$ can be effective in preventing destructive collisions between ultracold polar molecules. It is slightly less efficient than shielding on the transition $0\rightarrow 1$, but has some important advantages. In particular, it does not produce 2-molecule bound states under the conditions needed for shielding, so it will not enhance 3-body recombination. It thus obviates the need for double-field microwave shielding using a second field of different polarization.
\end{abstract}

\maketitle

\section{Introduction}

Ultracold polar molecules have emerged as a powerful and versatile platform for exploring diverse areas of quantum science. Their versatility arises from the strong and tunable dipolar interactions between molecules and their rich internal structure. These properties open the way to applications in quantum simulation \cite{Cornish:2024}, quantum magnetism \cite{Wall:QMUM:2015}, few- and many-body physics \cite{Greene:few-body:2017, Lahaye:2009, Baranov:2012}, and controlled chemical collisions \cite{Liu:bimol:2022, Karman:2024}.

A major obstacle in the production of ultracold molecular gases is the existence of strong collisional losses that occur at short range ($R \lesssim 100$ bohr). These losses arise from a variety of phenomena, including two-body inelastic collisions, three-body recombination \cite{Mayle:2013}, photoexcitation of collision complexes \cite{Christianen:laser:2019}, and chemical reactions \cite{Ospelkaus:react:2010, Hu:2019}. In molecular gases confined to two dimensions, such losses can be suppressed by applying a strong electric field perpendicular to the plane, which induces long-range repulsion between side-by-side dipoles \cite{deMiranda:2011, Valtolina:KRb2D:2020}. However, this technique is ineffective in three-dimensional geometries, where molecules can approach one another from arbitrary directions.

This problem may be overcome using collisional \emph{shielding} techniques based on static electric \cite{Avdeenkov:2006, Wang:dipolar:2015, Gonzalez-Martinez:adim:2017, Mukherjee:CaF:2023, Mukherjee:alkali:2024} or microwave fields \cite{Karman:shielding:2018, Lassabliere:2018, Karman:shielding-imp:2019, Karman:ellip:2020, Karman:res:2022, Deng:microwave:2023}. These techniques allow the production of stable molecular gases and provide exquisite control over their interactions. They rely on creating near-degeneracies between selected molecular pair states to engineer a long-range repulsive barrier that prevents molecules from undergoing destructive short-range collisions. They have allowed the creation of quantum molecular gases with high phase-space density \cite{Matsuda:2020, Li:KRb-shield-3D:2021, Schindewolf:NaK-degen:2022, Bigagli:NaCs:2023, Lin:NaRb:2023}. Shielding techniques have played a central role in achieving Fermi degeneracy \cite{Schindewolf:NaK-degen:2022} and Bose--Einstein condensation (BEC) \cite{Bigagli:BEC:NaCs:2024,Z_Shi:BEC:NaRb:2025} for molecules. These advances have established a versatile platform for exploring many-body physics with shielded polar molecules \cite{Langen:dipolar-droplets:2025, Jin:Bose:2025, Schindewolf:few-many:2025, Zhang:dipolar-droplets:2025, Zhang:supersolid:2025, Cardinale:supersolid:2025, Carroll:t-J:2025, Mukherjee:SU(N):2025}.

In both static-electric and microwave shielding schemes, the interactions between molecules can be characterized by anisotropic effective potentials \cite{Deng:microwave:2023, Deng:double-microwave:2025, Karman:double:2025, Mukherjee:eff-pot:2025}. These consist of a short-range repulsive core and a long-range attractive tail that together produce a well at very long range.  The qualitative features of these effective potentials are remarkably different for molecules shielded by a static electric field \cite{Mukherjee:eff-pot:2025} and by a single microwave field \cite{Deng:microwave:2023}. For static-field shielding, the long-range potential is dipolar in nature \cite{Bohn:BCT:2009}, whereas for single-field microwave shielding it has opposite sign, described as anti-dipolar \cite{Karman:res:2022}. This will lead, for example, to different geometries in the formation of self-bound droplets for these two types of shielded molecules. Another important difference is that microwave shielding, in contrast to static-field shielding, is universal in nature, offering very similar two-body collision properties for different molecules when expressed in suitable reduced units of length and energy \cite{Dutta:universal:2025}.

It is also useful to define one-dimensional effective potentials that take account of relative rotational motion \cite{Avdeenkov:2003}. For strongly dipolar molecules, such as NaRb, NaCs, and KAg, the long-range well produced by either type of shielding is sufficiently deep to host one or more two-molecule bound states \cite{Lassabliere:2018, Mukherjee:alkali:2024, Dutta:universal:2025}. The emergence of these bound states is characterized by poles in the scattering length as a function of field strength, which occur when the bound states cross the incoming threshold. The experimental realization of this two-molecule bound state has opened the door to creating ultracold polyatomic molecules \cite{Quemener:electroassoc:2023, Chen:field-linked-resonances:2023, Chen:field-linked-states:2024}. However, the presence of two-molecule bound states also enhances three-body losses in ultracold samples with high phase-space density and can prevent evaporative cooling to quantum degeneracy \cite{Stevenson:3body:2025}.

\begin{figure*}[tb]
\includegraphics[width=1.5\columnwidth]{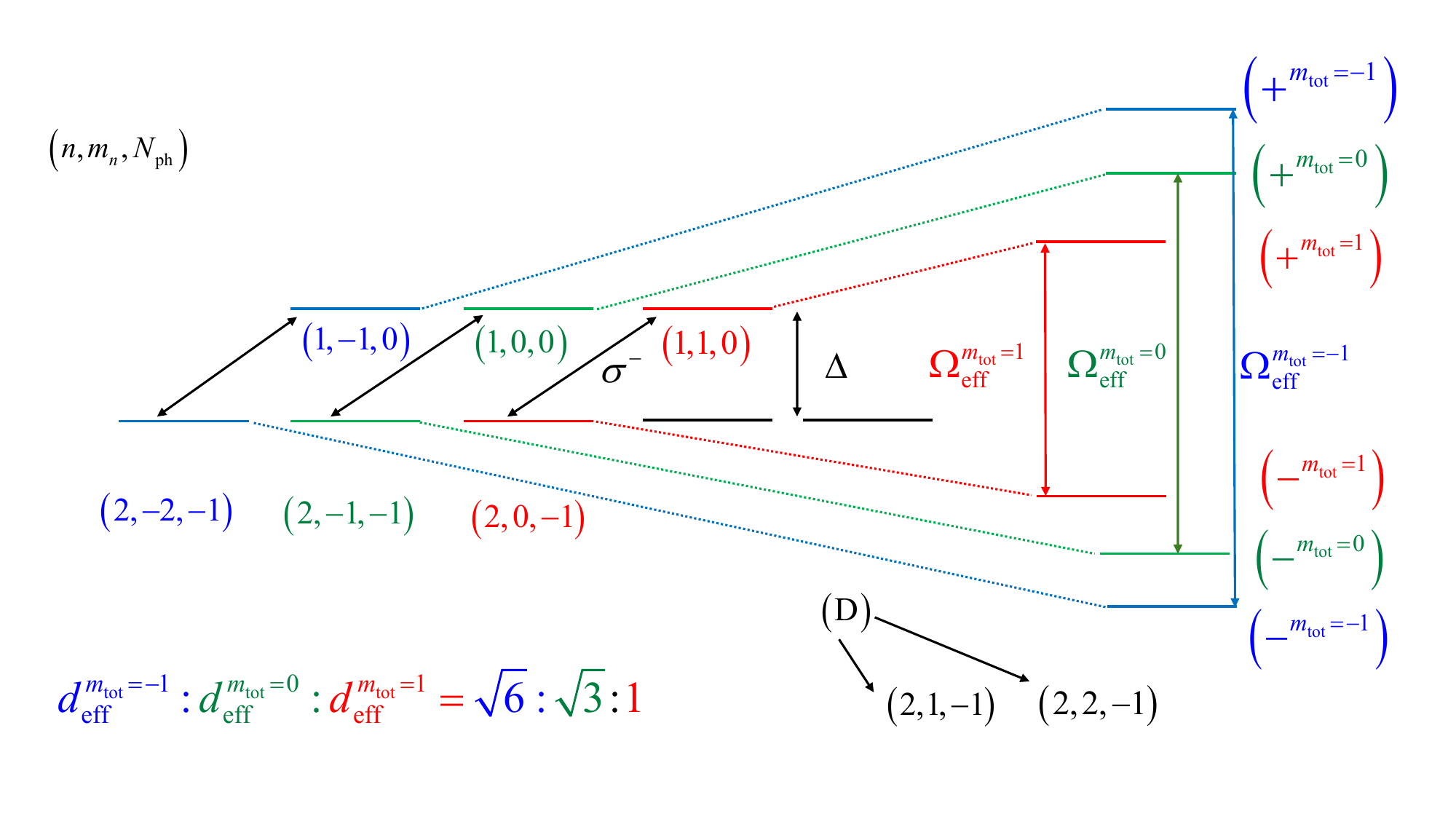}
\caption{Monomer interaction picture for microwave dressing on the transition $n=1 \rightarrow 2$, with blue-detuned microwave of polarization $\sigma^-$. There are three interacting blocks corresponding to three different values of $m_\textrm{tot}$.}
\label{fig:monomer-int}
\end{figure*}

The two-body bound states can be eliminated using double-field microwave shielding \cite{Bigagli:BEC:NaCs:2024, Karman:double:2025}. This prevents losses due to three-body recombination. It relies on adding a second microwave field of linear polarization to compensate the dipole-dipole interactions induced by circularly polarized microwaves. It played a pivotal role in the realization of BEC for NaCs \cite{Bigagli:BEC:NaCs:2024} and NaRb \cite{Z_Shi:BEC:NaRb:2025} molecules. Similar tunability of interactions between static-field shielded molecules can be achieved by adding a circularly polarized microwave field \cite{Ho:static-mw:2026, Wang:static-mw:2026}. These developments have set the stage for exploring novel, strongly correlated quantum phases with ultracold gases of dipolar molecules \cite{Schindewolf:few-many:2025, Zhang:dipolar-droplets:2025, Biswas:Fermi-surface-MW:2026}.

In this paper, we demonstrate that microwave shielding on the rotational transition $n=1\rightarrow2$ can be effective for typical strongly dipolar molecules, such as NaRb, NaCs, and KAg. Shielding without bound-state formation can be achieved with a circularly polarized microwave field alone, without requiring a second field of different polarization. With an appropriate choice of initial state, the loss rate can be suppressed to a level similar to that used to achieve BEC for NaCs and NaRb \cite{Bigagli:BEC:NaCs:2024, Z_Shi:BEC:NaRb:2025}.

\section{Theory}
\label{sec:theory}

The theory of microwave shielding has been given previously \cite{Karman:shielding:2018, Karman:shielding-imp:2019}, but will be summarized briefly here to define notation and point out ways in which shielding on the rotational transition $n=1\rightarrow 2$ differs from that on $0\rightarrow 1$.

\subsection{Interaction between molecule and microwave field}
\label{subsec:mol_mw}

The Hamiltonian that describes the molecule-microwave interaction is \citep{Cohen-Tannoudji:API:1998, Alyabyshev:microwave:2009}
\begin{equation}
\hat{h}_{\textrm{mw},\sigma} =   - \frac{F_\textrm{ac}}{2 \sqrt{N_0}} [\hat{\mu}_{\sigma} \hat{a}_{\sigma} + \hat{\mu}_{\sigma}^\dagger \hat{a}_{\sigma}^\dagger]
 + \hbar \omega (\hat{a}_{\sigma} \hat{a}^{\dagger}_{\sigma}  - N_0),
\label{eq:Hac}
\end{equation}
where $F_\textrm{ac}$ is the time-varying electric field and $N_0$ is the reference number of photons. Here, photons with polarization $\sigma$ are described by the creation and annihilation operators $\hat{a}^{\dagger}_{\sigma}$ and $\hat{a}_{\sigma}$, while the corresponding components of the dipole moment operator are $\hat{\mu}_{\sigma}^\dagger$ and $\hat{\mu}_{\sigma}$. In this paper we focus on shielding with circular polarization, $\sigma^-$, in the $xy$ plane. The microwave frequency $\omega$ is blue-detuned from the rotational transition $n=1 \rightarrow 2$ (at frequency $4b_\textrm{rot}/\hbar$) by $\Delta$, where $b_\textrm{rot}$ is the rotational constant of the molecule.

Figure \ref{fig:monomer-int} shows the monomer-microwave interaction diagram for the transition $n=1 \rightarrow 2$ with blue-detuned microwaves of polarization $\sigma^-$. The dressed levels are different for different values of $m_\textrm{tot} = m_n - N_\textrm{ph}$, where $m_n$ is the projection of the rotational quantum number $n$ onto $z$ and $N_\textrm{ph}$ is the number of photons with respect to the reference $N_0$. The monomer states $(n,m_n) = (2,1)$ and $(2,2)$ are dark states that are uncoupled from the $n = 1$ manifold and are denoted D. For a given value of $F_\textrm{ac}$, the Rabi frequency $\Omega$ and effective dipole moment $d_\textrm{eff}$ for the three blocks with $m_\textrm{tot}=-1$, 0 and $1$ are in the ratio $ \sqrt{6} : \sqrt{3} : 1$. The corresponding value for the transition $0\rightarrow 1$ is $\sqrt{5}$. For each value of $m_\textrm{tot}$, the splitting between the dressed states is the effective Rabi frequency $\Omega_\textrm{eff} = \sqrt{\Omega^2 + \Delta^2}$, which follows $\Omega_\textrm{eff}^{m_\textrm{tot}=-1} > \Omega_\textrm{eff}^{m_\textrm{tot}=0} > \Omega_\textrm{eff}^{m_\textrm{tot}=1}$.

\subsection{Coupled-channel formalism for microwave shielding}
\label{subsec:cc}

The Hamiltonian that describes a two-body collision in the center-of-mass frame is
\begin{equation}
\hat{H} = \frac{\hbar^2}{2\mu_\textrm{red}} \left ( - \frac{1}{R} \frac{d^{2}}{dR^{2}} R + \frac{\hat{\boldsymbol{L}}^2}{R^2} \right) + \hat{h}_\textrm{A} + \hat{h}_\textrm{B} + \hat{V}_\textrm{int}, \quad
\label{Eq: Dimer_Hamiltonian}
\end{equation}
where $\mu_\textrm{red}$ is the reduced mass of the colliding pair, $R$ is the intermolecular distance, and $\hat{\boldsymbol{L}}$ is the operator for end-over-end rotation of the pair. $\hat{h}_\textrm{A}$ and $\hat{h}_\textrm{B}$ are the internal Hamiltonians of the individual molecules within the rigid-rotor approximation, in the presence of a microwave radiation field as described in Eq.\ \ref{eq:Hac}, and $\hat{V}_\textrm{int}$ is the operator describing the interaction potential.

For shielding, the important interactions occur at long range, $R \gg 100$ bohr, so $\hat{V}_\textrm{int}$ is approximated by the dipole--dipole interaction,
\begin{equation}
\hat{H}_\textrm{dd} = -\frac{\sqrt{6}}{4 \pi \epsilon_0 R^3} T^{(2)}(\hat{\mu}_\textrm{A},\hat{\mu}_\textrm{B})
\cdot C^{(2)}(\boldsymbol{\hat{R}}),
\label{Eq: Dipole-Dipole Operator}
\end{equation}
where $T^{(2)}$ and $C^{(2)}$ are second-rank tensors, $\boldsymbol{\hat{R}}$ is a unit vector along the intermolecular axis and the components of $C^{(2)}(\boldsymbol{\hat{R}})$ are Racah-normalized spherical harmonics.  Contributions to $\hat{V}_\textrm{int}$ beyond the dipole--dipole interaction, such as higher-order multipole and dispersion terms, decay more rapidly with $R$ and play only a negligible role at shielding distances.

The total wavefunction is expanded
\begin{equation}
\Psi (R, \boldsymbol{\hat{R}}, \boldsymbol{\hat{r}}_\textrm{A}, \boldsymbol{\hat{r}}_\textrm{B}) = \frac{1}{R} \sum_{j} \psi_j (R) \Phi_{j} (\boldsymbol{\hat{R}}, \boldsymbol{\hat{r}}_\textrm{A}, \boldsymbol{\hat{r}}_\textrm{B}),
\label{Eq: Total Wavefunction}
\end{equation}
where $\boldsymbol{\hat{r}}_k$ is the unit vector along the axis of molecule $k$. The channel basis functions $\Phi_j$ are products of rotational pair-state functions, photon number states $N_\textrm{ph}$ (counted relative to $N_0$) and the
eigenfunctions of $\hat{\boldsymbol{L}}^2$, symmetrized as required for exchange of identical particles, as in Refs.\ \cite{Karman:shielding:2018, Dutta:universal:2025}.

Substituting the expansion (\ref{Eq: Total Wavefunction}) into the total Schr\"odinger equation yields a set coupled differential equations (the coupled-channel equations) in $R$. For pure $\sigma^-$ polarization, the projection $M_\textrm{tot} = m_{n_\textrm{A}}+m_{n_\textrm{B}}+M_L-N$ of the total angular momentum is a conserved quantity. We take advantage of this conservation by performing calculations for one value of $M_\textrm{tot}$ at a time.

The one-dimensional effective potentials (adiabats) $U_j(R)$ are defined as the eigenvalues of
\begin{equation}
\hat{h}_\textrm{A} + \hat{h}_\textrm{B} + \hat{V}_\textrm{int} + \frac{\hbar^2\hat{\boldsymbol{L}}^2}{2\mu_\textrm{red}R^2}
\end{equation}
at fixed $R$.

\subsection{Scattering observables: scattering length, cross section, and rate coefficients}
\label{subsec:scat_obs}

Shielding techniques can suppress collisional losses such that elastic rates exceed two-body loss rates by orders of magnitude. Elastic collisions follow the incoming adiabat, with the molecules remaining in the same states after the interaction. Two-body loss processes are primarily governed by nonadiabatic transitions to lower-lying channels. These transitions can lead either to short-range loss at small $R$ or to inelastic loss upon reflecting back to long range.

These processes are modeled by solving the coupled-channel equations with a fully absorbing boundary condition at short range \cite{Clary:1987, Janssen:PhD:2012, Mukherjee:CaF:2023}. The coupled-channel equations for microwave shielding are implemented as a plug-in basis-set suite for the MOLSCAT package \cite{molscat:2019, mbf-github:2025}. They are solved using the numerical methods described in Ref.\ \cite{Dutta:universal:2025}. This yields a nonunitary S matrix $\boldsymbol{S}$ for each set of microwave parameters ($\Omega$ and $\Delta/\Omega$) and collision energy $E_\textrm{coll}$. Cross sections for elastic scattering ($\sigma_\textrm{el}$), inelastic loss ($\sigma_\textrm{inel}$) and short-range loss ($\sigma_\textrm{sr}$) are calculated from the S-matrix elements as described in Ref.\ \cite{Mukherjee:CaF:2023}. The corresponding rate coefficients are $k=v\sigma$, where $v=(2E_\textrm{coll}/\mu_\textrm{red})^{1/2}$.  The sum of $k_\textrm{inel}$ and $k_\textrm{sr}$ is $k_\textrm{loss}$ and represents the total two-body loss.

We also calculate complex energy-dependent s-wave scattering lengths \citep{Hutson:res:2007},
\begin{eqnarray}
a(k_0) = \alpha(k_0)-i\beta(k_0) = \frac{1}{i k_0} \left( \dfrac{1-S_{00} (k_0)}{1+S_{00} (k_0)} \right),
\end{eqnarray}
where $k_0 = (2\mu_\textrm{red}E_\textrm{coll}/\hbar^2)^\frac{1}{2}$ is the incoming wavevector.

\begin{figure*}[tb]
    \centering
    \includegraphics[width=1.4\columnwidth]{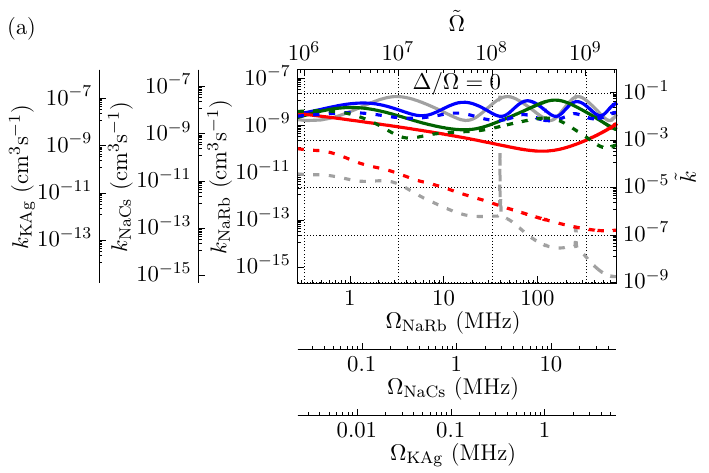}
    \hfill
    \includegraphics[width=1.4\columnwidth]{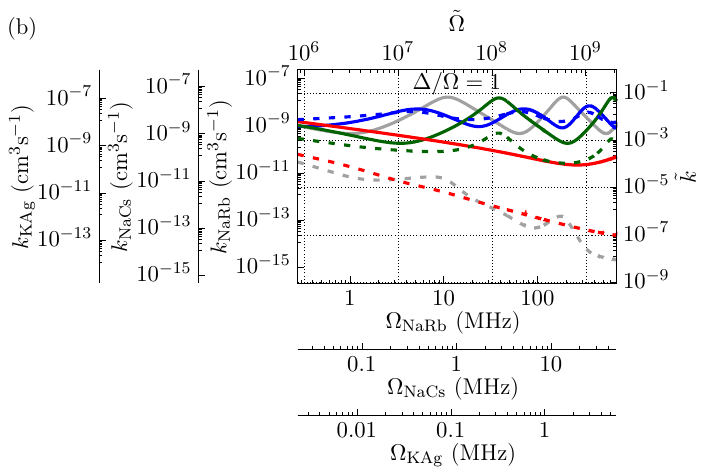}
    \caption{Rate coefficients for elastic scattering (solid lines) and total loss (dashed lines) as a function of $\Omega$ at (a) $\Delta/\Omega=0$ and (b) $\Delta/\Omega=1$ for microwave dressing on $n=0 \rightarrow 1$ (gray) and $n=1 \rightarrow 2$ for $(n,m_n)=(1,1)$ (red), $(1,0)$ (green) and $(1,-1)$ (blue). The calculations are performed at collision energy $E_\textrm{coll}=1000 E_3$. The values of $\Omega$ are defined separately for each transition, so the ac electric field $F_\textrm{ac}$ required to achieve a given $\Omega$ is different for each transition.}
    \label{fig:univ-rate-Rabi}
\end{figure*}

\subsection{Universality}

The coupled-channel equations take a dimensionless form when expressed in terms of reduced lengths $\tilde{R}=R/R_3$, reduced energies $\tilde{E}=E/E_3$, reduced Rabi frequencies $\tilde{\Omega}=\hbar\Omega/E_3$ and detunings $\tilde{\Delta}=\hbar\Delta/E_3$. Here $R_3=(2 \mu_\textrm{red}/ \hbar^2)(\mu^2/4 \pi \epsilon_0)$ and $E_3=\hbar^2/(2 \mu_\textrm{red} R_3^2)$ are the scaling factors for length and energy, respectively \cite{Gonzalez-Martinez:adim:2017}.
Reduced scattering properties are defined as $\tilde{\alpha} = \alpha/R_3$, $\tilde{\beta} = \beta/R_3$, $\tilde{\sigma}=\sigma/(4\pi R_3^2)$, and $\tilde{k}= k/k_3$, where $k_3= 4 \pi R_3 \hbar/\mu_\textrm{red}$.

The two-body collision properties of microwave-shielded molecules are universal \cite{Dutta:universal:2025}. When $\Omega, \Delta \ll b_\textrm{rot}/\hbar$, the shielding physics is dominated by channels within an energy window a few times $\hbar \Omega_\textrm{eff}$ wide around the incoming threshold. The energy separations between these near-resonant channels depend only on the microwave parameters $\Omega$ and $\Delta$. This produces a universal set of coupled-channel equations that are completely independent of molecular properties when expressed in the reduced units of length and energy. The one-dimensional effective potentials (adiabats) $U_j(R)$ are also identical for all molecules in these units. The resulting scattering observables, such as the scattering length, rate coefficients for elastic and inelastic scattering, and two-molecule bound states, are also universal. Deviations from universality are important only for very high Rabi frequencies.

For the transition $n=1\rightarrow 2$, the universal basis set involves 36 near-resonant pair states. This compares with 10 such states for universal calculations on the transition $n=0\rightarrow 1$ \cite{Dutta:universal:2025}. We carry out coupled-channel calculations using these basis sets. For each pair state, we include basis functions for the end-over-end angular momentum up to $L_\textrm{max}=12$, which ensures convergence of the rate coefficients to within 1\%. We neglect spin degrees of freedom, since their effects can be suppressed by a magnetic field sufficient to decouple the spins from the rotation \cite{Karman:shielding:2018}.

\section{Results and Discussion}
\label{sec:results}

In this section, we consider shielding on the transition $n=1 \rightarrow 2$ for three strongly dipolar molecules, NaRb, NaCs and KAg, which span a wide range of dipole moments. The scaling parameters for these molecules are given in Table \ref{tab:scaling_param}.

\subsection{Effectiveness of shielding}
\label{subsec:rate-alpha-vs-omg}

\begin{table}[tb]
\caption{Scaling and other parameters for the molecules considered in this study. The body-fixed dipole moments are taken from refs.\ \cite{Dagdigian:1972, Docenko:2006, Guo:NaRb:2016, Guo:NaRb:2018, Smialkowski:2021}.}
\label{tab:scaling_param}
\begin{ruledtabular}
\begin{tabular}{cccccc}
& $R_3$ (bohr)  & $E_3/k_\textrm{B}$ (K) & $k_3$ (cm$^3$ s$^{-1}$)  & $\mu$ (D)
 \\ \hline
$^{23}$Na$^{87}$Rb  &   $3.2 \times 10^5$ & $1.6 \times 10^{-11}$ & $2.4 \times 10^{-7}$ & 3.2  \\
$^{23}$Na$^{133}$Cs &   $9.9 \times 10^5$ & $1.1 \times 10^{-12}$ & $5.4 \times 10^{-7}$ &  4.75 \\
KAg\footnotemark\footnotetext[1]{Weighted-average atomic masses are used.} & $3.0 \times 10^6$ & $1.3 \times 10^{-13}$ & $1.7 \times 10^{-6}$  & 8.50 \\
\end{tabular}
\end{ruledtabular}
\end{table}

\begin{figure*}[tb]
    \centering
    \includegraphics[width=1.4\columnwidth]{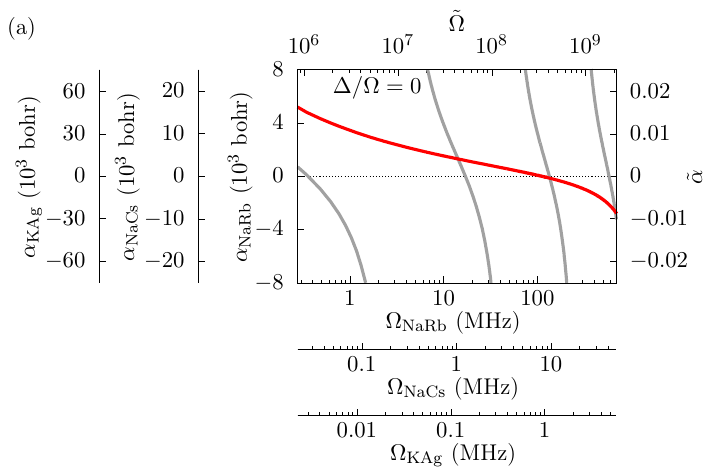}
    \hfill
    \includegraphics[width=1.4\columnwidth]{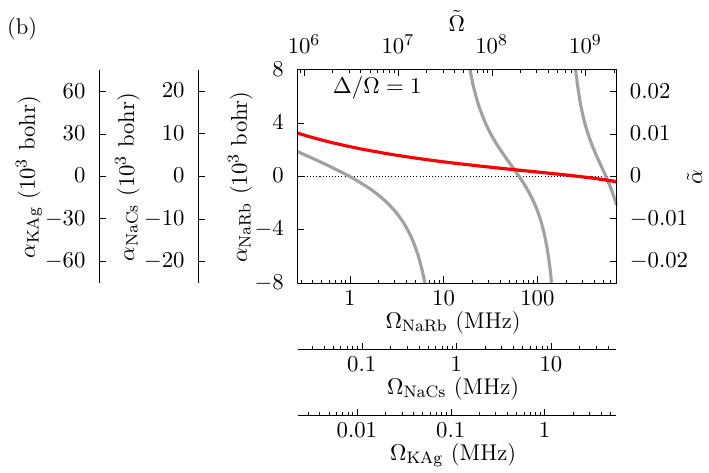}
\caption{Real part $\alpha$ of the scattering length as a function of $\Omega$ at (a) $\Delta/\Omega=0$ (b) $\Delta/\Omega=1$ for microwave dressing on $n=0 \rightarrow 1$ (gray) and $n=1 \rightarrow 2$ with $(n,m_n)=(1,1)$ (red). The calculations are performed at collision energy $E_\textrm{coll}=1000 E_3$. $\Omega$ shown on the x-axis is defined separately for each transition. Consequently, the corresponding ac electric field $F_\textrm{ac}$ required to achieve a given $\Omega$ differs for the three transitions.}
 \label{fig:univ-alpha-Rabi}
\end{figure*}

Figure \ref{fig:univ-rate-Rabi} shows rate coefficients with microwave shielding on the transition $n=1 \rightarrow 2$, with those for $n=0 \rightarrow 1$ included for comparison. The calculations are performed at collision energy $E_\textrm{coll}=1000 E_3$ for consistency with Ref.\ \cite{Dutta:universal:2025}.  The results for the three different incoming thresholds $(+^{m_\textrm{tot}} +^{m_\textrm{tot}})$ for $m_\textrm{tot}=-1$, 0 and 1 are shown in blue, green, and red, respectively. It is evident that shielding is most effective when molecules collide in the dressed state with $m_\textrm{tot}=1$, corresponding to molecules with predominant character $(n,m_n)=(1,1)$.

\begin{figure}[tb]
    \centering
    \includegraphics[width=1.0\columnwidth]{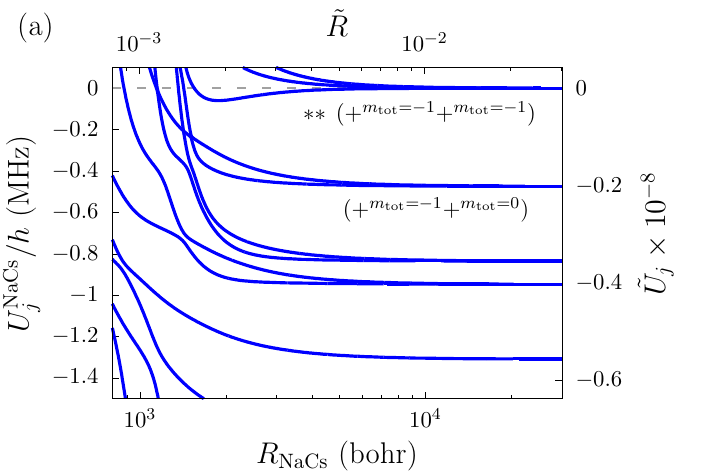}
    \hfill
    \includegraphics[width=1.0\columnwidth]{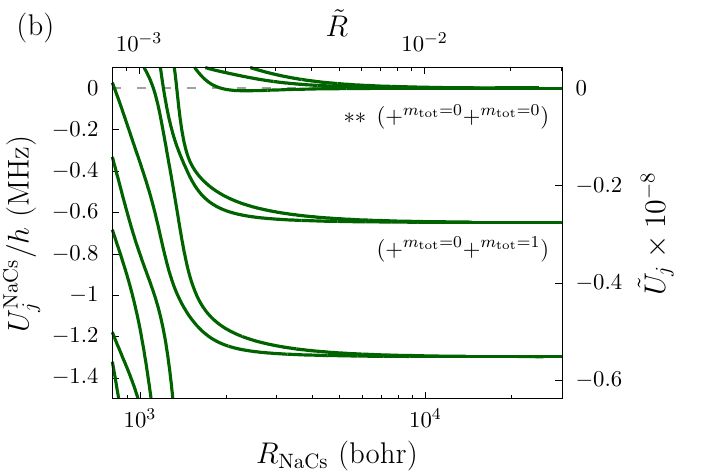}
    \hfill
    \includegraphics[width=1.0\columnwidth]{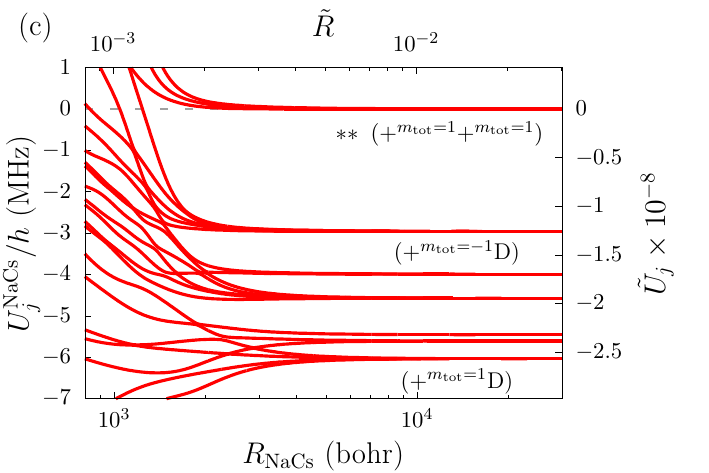}
\caption{Adiabats $U_j(R)$ for $L\le4$ for microwave shielding on the transition $n=1 \rightarrow 2$ for molecules that collide in the dressed state with (a) $m_\textrm{tot}=-1$ (blue), (b) $m_\textrm{tot}=0$ (green) and (c) $m_\textrm{tot}=1$ (red). The adiabats $U_j(R)$ and universal reduced adiabats $\tilde{U}_j(R)=U_j(R)/E_3$ are shown for $\tilde{\Omega}=2.1 \times 10^8$, corresponding to $\Omega=5$ MHz for NaCs, and $\Delta/\Omega=1$. The label $**$ indicates the incoming threshold for each $m_\textrm{tot}$.}
\label{fig:univ-adiabats}
\end{figure}

The slower losses for $m_\textrm{tot}=1$ arise from the energy separations between thresholds. Figure \ref{fig:univ-adiabats} shows the adiabats for the three values of $m_\textrm{tot}$. In each case, the dominant loss process involves nonadiabatic transitions from the incoming threshold $(+^{m_\textrm{tot}} +^{m_\textrm{tot}})$ to the threshold immediately below it.
For $m_\textrm{tot}=-1$ and 0, this involves a single-molecule transition to the next-lowest state. For $m_\textrm{tot}=1$, however, the single-molecule transition is further away and loss is dominated to transitions to other pair states in between. At $\Delta/\Omega=1$, this is the channel ($+^{m_\textrm{tot}=-1}$D), but even this has a substantially larger kinetic-energy release than for  $m_\textrm{tot}=-1$ and 0, and the difference increases with $\Delta/\Omega$. Nonadiabatic transitions are slower for larger kinetic energy releases, so are slowest for $m_\textrm{tot}=1$. The energy separations are discussed further in Sec.\ \ref{subsec:delta-omega} below.

\subsection{Two-molecule bound states}

Figure \ref{fig:univ-alpha-Rabi} compares the real part $\alpha$ of the scattering length for $n=1 \rightarrow 2$ (for $m_\textrm{tot}=1$) with that for $n=0 \rightarrow 1$. For the transition $n=0 \rightarrow 1$, the poles in $\alpha$ correspond to the appearance of two-molecule bound states. These also produce peaks in the rate coefficients for elastic scattering and loss, as shown in Fig.\ \ref{fig:univ-rate-Rabi}. For the transition $n=1 \rightarrow 2$, however, $\alpha$ exhibits no pole over the range in Fig.\ \ref{fig:univ-rate-Rabi}: there is no two-molecule bound state for $\tilde{\Omega}<2 \times 10^9 $ at any value of $\Delta/\Omega$. The range of $\tilde{\Omega}$ without bound states corresponds to at least $\Omega \sim 6$ MHz for KAg and $\Omega \sim 50$ MHz for NaCs.

\begin{figure*}[tb]
    \centering
    \includegraphics[width=1.0\columnwidth]{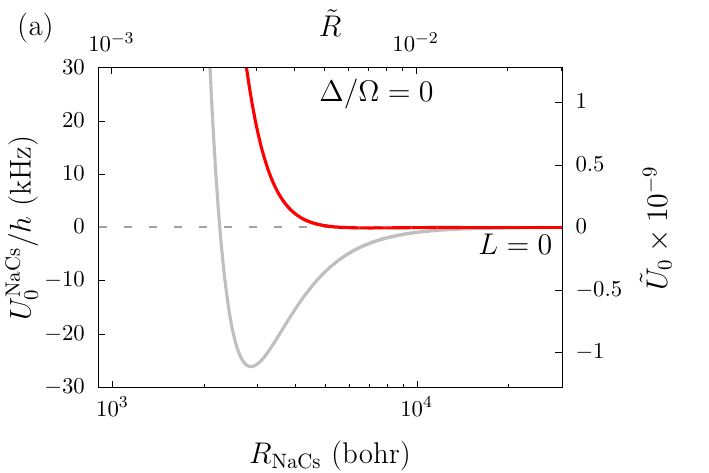}
    \hfill
    \includegraphics[width=1.0\columnwidth]{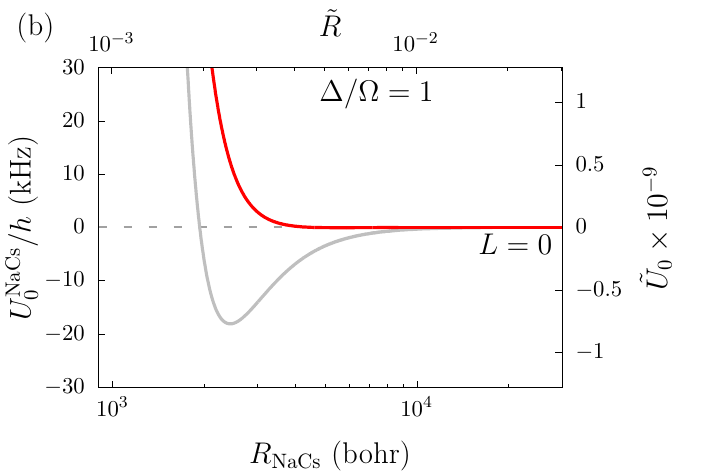}
\caption{Incoming s-wave adiabats $U_0 (R)$ for microwave dressing on $n=0 \rightarrow 1$ (gray) and $n=1 \rightarrow 2$ for $(1,1)$ (red) at (a) $\Delta/\Omega=0$ and (b) $\Delta/\Omega=1$ with fixed $\tilde{\Omega}=2.1 \times 10^8$, corresponding to $\Omega=5$ MHz for NaCs.}
\label{fig:univ-incoming-adiabat}
\end{figure*}

The absence of bound states for $n=1\rightarrow 2$ arises because the long-range attraction is much weaker than for $0\rightarrow 1$. The adiabats $U_0(R)$ for s-wave scattering in the two cases are compared in Fig.\ \ref{fig:univ-incoming-adiabat}. The attraction arises because $\hat{H}_\textrm{dd}$ mixes channels with $\Delta L=2$ that originate from the same threshold. $\hat{H}_\textrm{dd}$ is proportional to $R^{-3}$, while the energy separation between such channels scales as $R^{-2}$. This produces an effective long-range attraction of the form $-C_4/R^4$, with $C_4 = (2/15) (d_\textrm{eff}/\mu)^4 E_3 R_3^4$. Since $d_\textrm{eff}$ for shielding on $n=1\rightarrow 2$ with $m_\textrm{tot}=1$ is a factor of $\sqrt{5}$ times smaller than for $n=0 \rightarrow 1$, the dipole-dipole forces are a factor of 5 weaker and the resulting shielding well is much shallower.

\subsection{Choice of \protect{$\Delta/\Omega$}}
\label{subsec:delta-omega}

\begin{figure}[tb]
\includegraphics[width=1.0\columnwidth]{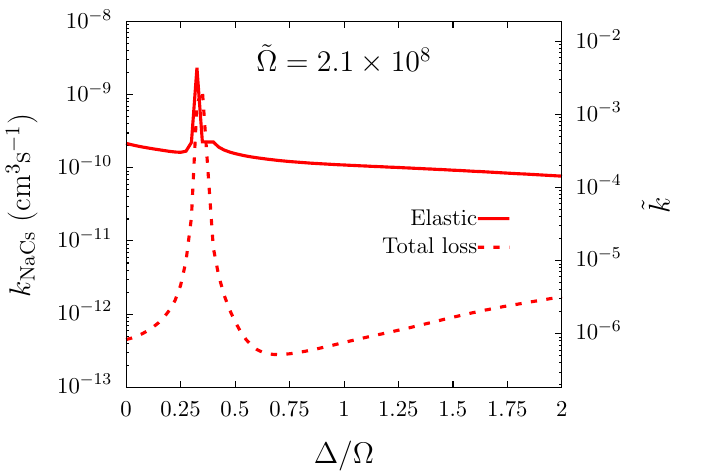}
\caption{Rate coefficients for elastic scattering (solid lines) and total loss (dashed lines) as a function of $\Delta/\Omega$ at $\tilde{\Omega}=2.1 \times 10^8$, corresponding to $\Omega=5$ MHz for NaCs, for microwave dressing on $n=1 \rightarrow 2$ for $(n,m_n)=(1,1)$. The calculations are performed at collision energy $E_\textrm{coll}=1000 E_3$.}
\label{fig:rate_vs_del_omg}
\end{figure}

\begin{figure}[tb]
\includegraphics[width=1.0\columnwidth]{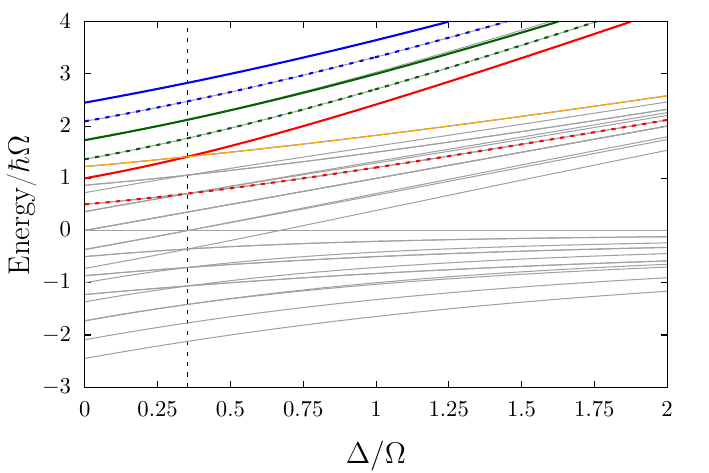}
\caption{Energies of field-dressed pair states as a function of $\Delta/\Omega$. The incoming thresholds $(++)$ for $m_\textrm{tot}=-1$, 0 and 1 are shown as solid blue, green, and red lines, respectively, and the single-molecule loss channel in each case is indicated with a corresponding dashed line. The threshold $(+^{m_\textrm{tot}=-1} \textrm{D})$ is shown in orange.}
\label{fig:pair-state-crossings}
\end{figure}

Figure \ref{fig:rate_vs_del_omg} shows the rate coefficients for elastic scattering and loss for $m_\textrm{tot}=1$ as a function of $\Delta/\Omega$. For illustration, we choose $\tilde{\Omega}=2.1 \times 10^8$, which corresponds to $\Omega=5$ MHz for NaCs. There is a strong peak in the loss rate coefficient near $\Delta/\Omega=0.35$; this arises from the crossing of the incoming threshold with the pair state $(+^{m_\textrm{tot}=-1}\textrm{D})$, where D is a dark state. The strongest suppression of loss occurs just above this peak, at $\Delta/\Omega \sim$ 0.70 for this value of $\tilde{\Omega}$.

Figure \ref{fig:pair-state-crossings} shows the energy of all the near-resonant pair states involved in microwave shielding for $n=1\rightarrow 2$. The energies are shown in units of $\Omega$, so apply to any value of $\Omega$. The incoming thresholds (++) for $m_\textrm{tot}=-1$, 0 and 1 are shown with colored solid lines, and the single-molecule loss channels are indicated with corresponding dashed lines. It may be seen that the incoming channels for shielding in both $m_\textrm{tot}=0$ and 1 cross other pair states at $\Delta/\Omega=0.3535$, and some other crossings between pair states occur at the same value of $\Delta/\Omega$. There is therefore unusually strong mixing between channels in this region, producing stronger loss.

The position of the loss maximum for $m_\textrm{tot}=1$ is almost independent of $\Omega$, but its shape is not. As $\tilde{\Omega}$ increases from the value chosen for Fig.\ \ref{fig:rate_vs_del_omg}, the position of the loss minimum shifts to slightly lower $\Delta/\Omega$ and the value of $\tilde{k}_\textrm{loss}$ at the minimum decreases.

\subsection{Extension to higher $n$}

We expect microwave shielding to be effective for higher rotational transitions, but to require stronger alternating fields $F_\textrm{ac}$. For each transition $n\rightarrow n+1$, the most effective component for shielding will usually be the one with the smallest intensity, since that is the one that places the single-molecule dressed state lowest in energy in the analog of Fig.\ \ref{fig:monomer-int}. This gives the largest energy separations for single-molecule nonadiabatic transitions and (in the absence of accidental near-degeneracies) is expected to give the lowest loss rates.

For shielding with blue-detuned microwaves with $\sigma^-$ polarization, the most favorable component of the transition $n\rightarrow n+1$ is $(n,m) = (n,n)\rightarrow (n+1,n-1)$. The transition moment for this component scales with $n$ as $[(2n+1)(2n+3)]^{-1/2}$. The effective dipole moment $d_\textrm{eff}$ scales by the same factor. The value of $F_\textrm{ac}$ needed to achieve the same Rabi frequency thus scales as $[(2n+1)(2n+3)]^{1/2}$. The dipole-dipole interactions scale with $1/[(2n+1)(2n+3)]$.

\section{Conclusions}
\label{sec:conc}

We have shown that microwave shielding on the rotational transition $n=1\rightarrow2$ can be effective at suppressing destructive collisions that occur during close collisions. For a microwave field of $\sigma^-$ polarization, it is most effective when the molecules collide in a microwave-dressed state with predominant character $(n,m_n)=(1,1)$. Collisional loss is slowest for this state because the energy gaps to lower states are larger than for $m_n=0$ or $-1$. The loss rate coefficient can be suppressed to levels comparable to those achieved in recent experiments that reached molecular BEC \cite{Bigagli:BEC:NaCs:2024, Z_Shi:BEC:NaRb:2025}.

The effective dipole moment for the shielded state is a factor of $\sqrt{5}$ smaller than for shielding on the transition $n=0\rightarrow 1$. Because of this, the interaction potential has a much shallower long-range well, which does not support a 2-molecule bound state except at very high Rabi frequencies. As a result, 3-body recombination cannot occur under the conditions required for shielding. This obviates the need for double-field microwave shielding, using a second microwave field with different polarization.

\section*{Rights retention statement}

For the purpose of open access, the authors have applied a Creative Commons Attribution (CC BY) licence to any Author Accepted Manuscript version arising from this submission.

\section*{Data availability statement}

The data that support the findings of this article
will be openly available. Please contact the authors for URL.

\section*{Acknowledgement}
We are grateful to Simon Cornish, Philip Gregory and Ruth Le Sueur for valuable discussions.
This work was supported by the U.K. Engineering and Physical Sciences Research Council (EPSRC) Grant Nos.\ 
EP/W00299X/1, 
EP/Z535898/1, 
and UKRI2226. 

\bibliography{../all}

\begin{thebibliography}{10}
\newcommand{\enquote}[1]{``#1''}

\bibitem{Cornish:2024}
S.~L. Cornish, M.~R. Tarbutt, and K.~R.~A. Hazzard.
\newblock \enquote{Quantum computation and quantum simulation with ultracold
  molecules.}
\newblock Nat. Phys., \textbf{20}, 730 (2024).

\bibitem{Wall:QMUM:2015}
M.~L. Wall, K.~R.~A. Hazzard, and A.~M. Rey.
\newblock \enquote{Quantum magnetism with ultracold molecules.}
\newblock In S.~A. Malinovskaya and I.~Novikova (Editors), \enquote{From Atomic
  to Mesoscale,}  (World Scientific, Singapore, 2015), chapter~1, 3--37.

\bibitem{Greene:few-body:2017}
C.~H. Greene, P.~Giannakeas, and J.~P\'erez-R\'{\i}os.
\newblock \enquote{Universal few-body physics and cluster formation.}
\newblock Rev. Mod. Phys., \textbf{89}, 035006 (2017).

\bibitem{Lahaye:2009}
T.~Lahaye, C.~Menotti, L.~Santos, M.~Lewenstein, and T.~Pfau.
\newblock \enquote{The physics of dipolar bosonic quantum gases.}
\newblock Rep. Prog. Phys., \textbf{72}, 126401 (2009).

\bibitem{Baranov:2012}
M.~A. Baranov, M.~Dalmonte, G.~Pupillo, and P.~Zoller.
\newblock \enquote{Condensed matter theory of dipolar quantum gases.}
\newblock Chem. Rev., \textbf{112}, 5012 (2012).

\bibitem{Liu:bimol:2022}
Y.~Liu and K.-K. Ni.
\newblock \enquote{Bimolecular chemistry in the ultracold regime.}
\newblock Ann. Rev. Phys. Chem., \textbf{73}, 73 (2022).

\bibitem{Karman:2024}
T.~Karman, M.~Tomza, and J.~P\'erez-R\'ioz.
\newblock \enquote{Ultracold chemistry as a testbed for few-body physics.}
\newblock Nature Physics, \textbf{20}, 722 (2024).

\bibitem{Mayle:2013}
M.~Mayle, G.~Qu\'em\'ener, B.~P. Ruzic, and J.~L. Bohn.
\newblock \enquote{Scattering of ultracold molecules in the highly resonant
  regime.}
\newblock Phys. Rev. A, \textbf{87}, 012709 (2013).

\bibitem{Christianen:laser:2019}
A.~Christianen, M.~W. Zwierlein, G.~C. Groenenboom, and T.~Karman.
\newblock \enquote{Photoinduced two-body loss of ultracold molecules.}
\newblock Phys. Rev. Lett., \textbf{123}, 123402 (2019).

\bibitem{Ospelkaus:react:2010}
S.~Ospelkaus, K.-K. Ni, D.~Wang, M.~H.~G. {de Miranda}, B.~Neyenhuis,
  G.~Qu\'{e}m\'{e}ner, P.~S. Julienne, J.~L. Bohn, D.~S. Jin, and J.~Ye.
\newblock \enquote{Quantum-state controlled chemical reactions of ultracold
  {KRb} molecules.}
\newblock Science, \textbf{327}, 853 (2010).

\bibitem{Hu:2019}
M.-G. Hu, Y.~Liu, D.~D. Grimes, Y.-W. Lin, A.~H. Gheorghe, R.~Vexiau,
  N.~Bouloufa-Maafa, O.~Dulieu, T.~Rosenband, and K.-K. Ni.
\newblock \enquote{Direct observation of bimolecular reactions of ultracold
  {KRb} molecules.}
\newblock Science, \textbf{366}, 1111 (2019).

\bibitem{deMiranda:2011}
M.~H.~G. de~Miranda, A.~Chotia, B.~Neyenhuis, D.~Wang, G.~Qu\'em\'ener,
  S.~Ospelkaus, J.~L. Bohn, J.~Ye, and D.~S. Jin.
\newblock \enquote{Controlling the quantum stereodynamics of ultracold
  bimolecular reactions.}
\newblock Nat. Phys., \textbf{7}, 502 (2011).

\bibitem{Valtolina:KRb2D:2020}
G.~Valtolina, K.~Matsuda, W.~G. Tobias, J.-R. Li, L.~De~Marco, and J.~Ye.
\newblock \enquote{Dipolar evaporation of reactive molecules to below the
  {F}ermi temperature.}
\newblock Nature, \textbf{588}, 239 (2020).

\bibitem{Avdeenkov:2006}
A.~V. Avdeenkov, M.~Kajita, and J.~L. Bohn.
\newblock \enquote{Suppression of inelastic collisions of polar {$^1\Sigma$}
  state molecules in an electrostatic field.}
\newblock Phys. Rev. A, \textbf{73}, 022707 (2006).

\bibitem{Wang:dipolar:2015}
G.~Wang and G.~Qu\'em\'ener.
\newblock \enquote{Tuning ultracold collisions of excited rotational dipolar
  molecules.}
\newblock New J. Phys., \textbf{17}, 035015 (2015).

\bibitem{Gonzalez-Martinez:adim:2017}
M.~L. Gonz\'{a}lez-Mart\'{\i}nez, J.~L. Bohn, and G.~Qu\'em\'ener.
\newblock \enquote{Adimensional theory of shielding in ultracold collisions of
  dipolar rotors.}
\newblock Phys. Rev. A, \textbf{96}, 032718 (2017).

\bibitem{Mukherjee:CaF:2023}
B.~Mukherjee, M.~D. Frye, C.~R. {Le Sueur}, M.~R. Tarbutt, and J.~M. Hutson.
\newblock \enquote{Shielding collisions of ultracold {CaF} molecules with
  static electric fields.}
\newblock Phys. Rev. Res., \textbf{5}, 033097 (2023).

\bibitem{Mukherjee:alkali:2024}
B.~Mukherjee and J.~M. Hutson.
\newblock \enquote{Controlling collisional loss and scattering lengths of
  ultracold dipolar molecules with static electric fields.}
\newblock Phys. Rev. Res., \textbf{6}, 013145 (2024).

\bibitem{Karman:shielding:2018}
T.~Karman and J.~M. Hutson.
\newblock \enquote{Microwave shielding of ultracold polar molecules.}
\newblock Phys. Rev. Lett., \textbf{121}, 163401 (2018).

\bibitem{Lassabliere:2018}
L.~Lassabli{\`e}re and G.~Qu{\'e}m{\'e}ner.
\newblock \enquote{Controlling the scattering length of ultracold dipolar
  molecules.}
\newblock Phys. Rev. Lett., \textbf{121}, 163402 (2018).

\bibitem{Karman:shielding-imp:2019}
T.~Karman and J.~M. Hutson.
\newblock \enquote{Microwave shielding of ultracold polar molecules with
  imperfectly circular polarization.}
\newblock Phys. Rev. A, \textbf{100}, 052704 (2019).

\bibitem{Karman:ellip:2020}
T.~Karman.
\newblock \enquote{Microwave shielding with far-from-circular polarization.}
\newblock Phys. Rev. A, \textbf{101}, 042702 (2020).

\bibitem{Karman:res:2022}
T.~Karman, Z.~Z. Yan, and M.~Zwierlein.
\newblock \enquote{Resonant and first-order dipolar interactions between
  ultracold {$^1\Sigma$} molecules in static and microwave electric fields.}
\newblock Phys. Rev. A, \textbf{105}, 013321 (2022).

\bibitem{Deng:microwave:2023}
F.~Deng, X.-Y. Chen, X.-Y. Luo, W.~Zhang, S.~Yi, and T.~Shi.
\newblock \enquote{Effective potential and superfluidity of microwave-shielded
  polar molecules.}
\newblock Phys. Rev. Lett., \textbf{130}, 183001 (2023).

\bibitem{Matsuda:2020}
K.~Matsuda, L.~De~Marco, J.-R. Li, W.~G. Tobias, G.~Valtolina, G.~Qu\'em\'ener,
  and J.~Ye.
\newblock \enquote{Resonant collisional shielding of reactive molecules using
  electric fields.}
\newblock Science, \textbf{370}, 1324 (2020).

\bibitem{Li:KRb-shield-3D:2021}
J.-R. Li, W.~G. Tobias, K.~Matsuda, C.~Miller, G.~Valtolina, L.~De~Marco,
  R.~R.~W. Wang, L.~Lassabli\`ere, G.~Qu\'em\'ener, J.~L. Bohn, and J.~Ye.
\newblock \enquote{Tuning of dipolar interactions and evaporative cooling in a
  three-dimensional molecular quantum gas.}
\newblock Nat. Phys., \textbf{17}, 1144 (2021).

\bibitem{Schindewolf:NaK-degen:2022}
A.~Schindewolf, R.~Bause, X.-Y. Chen, M.~Duda, T.~Karman, I.~Bloch, and X.-Y.
  Luo.
\newblock \enquote{Evaporation of microwave-shielded polar molecules to quantum
  degeneracy.}
\newblock Nature, \textbf{607}, 677 (2022).

\bibitem{Bigagli:NaCs:2023}
N.~Bigagli, C.~Warner, W.~Yuan, S.~Zhang, I.~Stevenson, T.~Karman, and S.~Will.
\newblock \enquote{Collisionally stable gas of bosonic dipolar ground-state
  molecules.}
\newblock Nat. Phys., \textbf{19}, 1579 (2023).

\bibitem{Lin:NaRb:2023}
J.~Lin, G.~Chen, M.~Jin, Z.~Shi, F.~Deng, W.~Zhang, G.~Qu\'em\'ener, T.~Shi,
  S.~Yi, and D.~Wang.
\newblock \enquote{Microwave shielding of bosonic {NaRb} molecules.}
\newblock Phys. Rev. X, \textbf{13}, 031032 (2023).

\bibitem{Bigagli:BEC:NaCs:2024}
N.~Bigagli, W.~Yuan, S.~Zhang, B.~Bulatovic, T.~Karman, I.~Stevenson, and
  S.~Will.
\newblock \enquote{Observation of {Bose-Einstein} condensation of dipolar
  molecules.}
\newblock Nature, \textbf{631}, 289 (2024).

\bibitem{Z_Shi:BEC:NaRb:2025}
Z.~Shi, Z.~Huang, F.~Deng, W.-J. Jin, S.~Yi, T.~Shi, and D.~Wang.
\newblock \enquote{Bose-{E}instein condensate of ultracold sodium-rubidium
  molecules with tunable dipolar interactions.}
\newblock arXiv:2508.20518 (2025).

\bibitem{Langen:dipolar-droplets:2025}
T.~Langen, J.~Boronat, J.~S\'anchez-Baena, R.~Bomb\'in, T.~Karman, and
  F.~Mazzanti.
\newblock \enquote{Dipolar droplets of strongly interacting molecules.}
\newblock Phys. Rev. Lett., \textbf{134}, 053001 (2025).

\bibitem{Jin:Bose:2025}
W.-J. Jin, F.~Deng, S.~Yi, and T.~Shi.
\newblock \enquote{Bose-{E}instein condensates of microwave-shielded polar
  molecules.}
\newblock Phys. Rev. Lett., \textbf{134}, 233003 (2025).

\bibitem{Schindewolf:few-many:2025}
A.~Schindewolf, J.~Hertkorn, I.~Stevenson, M.~Ciardi, P.~Gross, D.~Wang,
  T.~Karman, G.~Qu\'em\'ener, S.~Will, T.~Pohl, and T.~Langen.
\newblock \enquote{From few- to many-body physics: Strongly dipolar molecular
  {B}ose-{E}instein condensates and quantum fluids.}
\newblock arXiv:2512.14511 (2025).

\bibitem{Zhang:dipolar-droplets:2025}
S.~Zhang, W.~Yuan, N.~Bigagli, H.~Kwak, T.~Karman, I.~Stevenson, and S.~Will.
\newblock \enquote{Observation of self-bound droplets of ultracold dipolar
  molecules.}
\newblock arXiv:2507.15208 (2025).

\bibitem{Zhang:supersolid:2025}
W.~Zhang, H.~Liu, F.~Deng, K.~Chen, S.~Yi, and T.~Shi.
\newblock \enquote{Supersolid phases in ultracold gases of microwave shielded
  polar molecules.}
\newblock arXiv:2506.23820 (2025).

\bibitem{Cardinale:supersolid:2025}
T.~A. Cardinale, T.~Bland, and S.~M. Reimann.
\newblock \enquote{Exploring molecular supersolidity via exact and mean-field
  theories: single microwave shielding.}
\newblock arXiv:2509.18051 (2025).

\bibitem{Carroll:t-J:2025}
A.~N. Carroll, H.~Hirzler, C.~Miller, D.~Wellnitz, S.~R. Muleady, J.~Lin, K.~P.
  Zamarski, R.~R.~W. Wang, J.~L. Bohn, A.~M. Rey, and J.~Ye.
\newblock \enquote{Observation of generalized $t$-${J}$ spin dynamics with
  tunable dipolar interactions.}
\newblock Science, \textbf{388}, 381 (2025).

\bibitem{Mukherjee:SU(N):2025}
B.~Mukherjee, J.~M. Hutson, and K.~R.~A. Hazzard.
\newblock \enquote{{SU($N$)} magnetism with ultracold molecules.}
\newblock New J. Phys., \textbf{27}, 013013 (2025).

\bibitem{Deng:double-microwave:2025}
F.~Deng, X.~Hu, W.-J. Jin, S.~Yi, and T.~Shi.
\newblock \enquote{Two- and many-body physics of ultracold molecules dressed by
  dual microwave fields.}
\newblock Nat. Commun., \textbf{16}, 11219 (2025).

\bibitem{Karman:double:2025}
T.~Karman, N.~Bigagli, W.~Yuan, S.~Zhang, I.~Stevenson, and S.~Will.
\newblock \enquote{Double microwave shielding.}
\newblock PRX Quantum, \textbf{6}, 020358 (2025).

\bibitem{Mukherjee:eff-pot:2025}
B.~Mukherjee, L.~Santos, and J.~M. Hutson.
\newblock \enquote{Effective anisotropic interaction potentials for pairs of
  ultracold molecules shielded by a static electric field.}
\newblock New J. Phys., \textbf{27}, 093204 (2025).

\bibitem{Bohn:BCT:2009}
J.~L. Bohn, M.~Cavagnero, and C.~Ticknor.
\newblock \enquote{Quasi-universal dipolar scattering in cold and ultracold
  gases.}
\newblock New J. Phys., \textbf{11}, 055039 (2009).

\bibitem{Dutta:universal:2025}
J.~Dutta, B.~Mukherjee, and J.~M. Hutson.
\newblock \enquote{Universality in the microwave shielding of ultracold polar
  molecules.}
\newblock Phys. Rev. Res, \textbf{7}, 023164 (2025).

\bibitem{Avdeenkov:2003}
A.~V. Avdeenkov and J.~L. Bohn.
\newblock \enquote{Linking ultracold polar molecules.}
\newblock Phys. Rev. Lett., \textbf{90}, 043006 (2003).

\bibitem{Quemener:electroassoc:2023}
G.~Qu\'em\'ener, J.~L. Bohn, and J.~F.~E. Croft.
\newblock \enquote{Electroassociation of ultracold dipolar molecules into
  tetramer field-linked states.}
\newblock Phys. Rev. Lett., \textbf{131}, 043402 (2023).

\bibitem{Chen:field-linked-resonances:2023}
X.-Y. Chen, A.~Schindewolf, S.~Eppelt, R.~Bause, M.~Duda, S.~Biswas, T.~Karman,
  T.~Hilker, I.~Bloch, and X.-Y. Luo.
\newblock \enquote{Field-linked resonances of polar molecules.}
\newblock Nature, \textbf{614}, 59 (2023).

\bibitem{Chen:field-linked-states:2024}
X.-Y. Chen, S.~Biswas, S.~Eppelt, A.~Schindewolf, F.~Deng, T.~Shi, S.~Yi, T.~A.
  Hilker, I.~Bloch, and X.-Y. Luo.
\newblock \enquote{Ultracold field-linked tetratomic molecules.}
\newblock Nature, \textbf{626}, 283 (2024).

\bibitem{Stevenson:3body:2025}
I.~Stevenson, S.~Singh, A.~Elkamshishy, N.~Bigagli, W.~Yuan, S.~Zhang, C.~H.
  Greene, and S.~Will.
\newblock \enquote{Three-body recombination of ultracold microwave-shielded
  polar molecules.}
\newblock Phys. Rev. Lett., \textbf{133}, 263402 (2024).

\bibitem{Ho:static-mw:2026}
C.~J. Ho, J.~Dutta, B.~Mukherjee, J.~M. Hutson, and M.~R. Tarbutt.
\newblock \enquote{Tuning interactions between static-field-shielded polar
  molecules with microwaves.}
\newblock Phys. Rev. Res., \textbf{8}, 023087 (2026).

\bibitem{Wang:static-mw:2026}
R.~R.~W. Wang.
\newblock \enquote{Bound-state-free {F}\"orster resonant shielding of strongly
  dipolar ultracold molecules.}
\newblock arXiv:2601.21928 (2026).

\bibitem{Biswas:Fermi-surface-MW:2026}
S.~Biswas, S.~Eppelt, W.~Tian, W.~Zhang, F.~Deng, C.~Frank, T.~Shi, I.~Bloch,
  and X.-Y. Luo.
\newblock \enquote{Controlled symmetry breaking of the {F}ermi surface in
  ultracold polar molecules.} (2026).

\bibitem{Cohen-Tannoudji:API:1998}
C.~Cohen-Tannoudji, J.~Dupont-Roc, and G.~Grynberg.
\newblock \emph{Atom-Photon Interactions: Basic Processes and Applications}
  (Wiley, New York, 1998).

\bibitem{Alyabyshev:microwave:2009}
S.~V. Alyabyshev and R.~V. Krems.
\newblock \enquote{Controlling collisional spin relaxation of cold molecules
  with microwave laser fields.}
\newblock Phys. Rev. A, \textbf{80}, 033419 (2009).

\bibitem{Clary:1987}
D.~C. Clary and J.~P. Henshaw.
\newblock \enquote{Chemical reactions dominated by long-range intermolecular
  forces.}
\newblock Faraday Discuss. Chem. Soc., \textbf{84}, 333 (1987).

\bibitem{Janssen:PhD:2012}
L.~M.~C. Janssen.
\newblock \emph{Cold collision dynamics of {NH} radicals}.
\newblock Ph.D. thesis, Radboud University, Nijmegen (2012).

\bibitem{molscat:2019}
J.~M. Hutson and C.~R. Le~Sueur.
\newblock \enquote{{\sc molscat}: a program for non-reactive quantum scattering
  calculations on atomic and molecular collisions.}
\newblock Comp. Phys. Comm., \textbf{241}, 9 (2019).

\bibitem{mbf-github:2025}
J.~M. Hutson and C.~R. Le~Sueur.
\newblock \enquote{{\sc molscat}, {\sc bound} and {\sc field}, version 2025.0.}
\newblock \url{https://github.com/molscat/molscat} (2025).

\bibitem{Hutson:res:2007}
J.~M. Hutson.
\newblock \enquote{Feshbach resonances in ultracold atomic and molecular
  collisions: threshold behaviour and suppression of poles in scattering
  lengths.}
\newblock New J. Phys., \textbf{9}, 152 (2007).

\bibitem{Dagdigian:1972}
P.~J. Dagdigian and L.~Wharton.
\newblock \enquote{Molecular beam electric deflection and resonance
  spectroscopy of the heteronuclear alkali dimers: {$^{39}$K$^7$Li},
  {Rb$^7$Li}, {$^{39}$K$^{23}$Na}, {Rb$^{23}$Na}, and {$^{133}$Cs$^{23}$Na}.}
\newblock J.~Chem. Phys., \textbf{57}, 1487 (1972).

\bibitem{Docenko:2006}
O.~Docenko, M.~Tamanis, J.~Zaharova, R.~Ferber, A.~Pashov, H.~Kn\"ockel, and
  E.~Tiemann.
\newblock \enquote{The coupling of the {X$^1\Sigma^+$} and {a$^3\Sigma^+$}
  states of the atom pair {Na} + {Cs} and modelling cold collisions.}
\newblock J. Phys. B - At. Mol. Opt., \textbf{39}, S929 (2006).

\bibitem{Guo:NaRb:2016}
M.~Guo, B.~Zhu, B.~Lu, X.~Ye, F.~Wang, R.~Vexiau, N.~Bouloufa-Maafa,
  G.~Qu\'em\'ener, O.~Dulieu, and D.~Wang.
\newblock \enquote{Creation of an ultracold gas of ground-state dipolar
  $^{23}${Na}$^{87}${Rb} molecules.}
\newblock Phys. Rev. Lett., \textbf{116}, 205303 (2016).

\bibitem{Guo:NaRb:2018}
M.~Guo, X.~Ye, J.~He, G.~Qu{\'e}m{\'e}ner, and D.~Wang.
\newblock \enquote{High-resolution internal state control of ultracold
  {$^{23}$Na$^{87}$Rb} molecules.}
\newblock Phys. Rev. A, \textbf{97}, 020501(R) (2018).

\bibitem{Smialkowski:2021}
M.~\'{S}mia\l{}kowski and M.~Tomza.
\newblock \enquote{Highly polar molecules consisting of a copper or silver atom
  interacting with an alkali-metal or alkaline-earth-metal atom.}
\newblock Phys. Rev. A, \textbf{103}, 022802 (2021).

\end{thebibliography}
\bibliographystyle{long_bib}
\end{document}